\documentclass{vgtc}                          

\graphicspath{{figures/}{pictures/}{images/}{./}}

\usepackage{times}

\usepackage{tabu}
\usepackage{booktabs}
\usepackage{tabularx}
\usepackage{lipsum}
\usepackage{mwe}
\usepackage{xcolor}
\usepackage{colortbl}
\usepackage{multirow}
\usepackage{rotating}
\usepackage{pdflscape}
\usepackage{amsmath}
\usepackage{mathptmx}

\DeclareRobustCommand{\rr}[1]{#1}

\usepackage{longtable}
\usepackage{array,ragged2e,tabularx,booktabs}
\usepackage[most]{tcolorbox}
\usepackage{xstring}
\usepackage{pgf}
\usepackage[table,xcdraw]{xcolor}
\usepackage{xspace}
\usepackage{float}

\renewcommand{\arraystretch}{1.15}
\newcolumntype{Y}{>{\RaggedRight\arraybackslash\hspace{0pt}\vspace{0pt}}X}

\newcommand{\codewrap}[1]{\ttfamily\RaggedRight\hspace{0pt}#1}

\definecolor{ScoreBlue}{HTML}{1170AA}
\definecolor{ScoreYellow}{HTML}{F2B134}
\definecolor{ScoreRed}{HTML}{E54D37}
\colorlet{ScoreBlueBG}{ScoreBlue!10}
\colorlet{ScoreYellowBG}{ScoreYellow!10}
\colorlet{ScoreRedBG}{ScoreRed!10}

\newcommand{\makeScoreBadge}[3]{%
  \raisebox{-0.2ex}{%
    \tcbox[
      on line,
      colback=#2, colframe=#2,
      boxrule=0pt, arc=2pt,
      left=2pt, right=2pt, top=0.2ex, bottom=0.2ex,
      boxsep=0pt,
      nobeforeafter
    ]{\textcolor{#1}{\textbf{#3}}}%
  }%
}

\newcommand{\scorebadge}[1]{%
  \begingroup
    \def\raw{#1}%
    \IfEndWith{\raw}{\%}{\StrGobbleRight{\raw}{1}[\rawnum]}{\def\rawnum{\raw}}%
    \pgfmathparse{\rawnum}\let\num\pgfmathresult
    \def\fg{ScoreBlue}\def\bg{ScoreBlueBG}%
    \pgfmathparse{\num<33?1:0}\ifnum\pgfmathresult=1
      \def\fg{ScoreRed}\def\bg{ScoreRedBG}%
    \else
      \pgfmathparse{\num<=65?1:0}\ifnum\pgfmathresult=1
        \def\fg{ScoreYellow}\def\bg{ScoreYellowBG}%
      \fi
    \fi
    \makeScoreBadge{\fg}{\bg}{#1}%
  \endgroup
}

\usepackage{xspace}
\newcommand{\lexararf}{\textsc{Lexara-RF}\xspace}

\newenvironment{tight_itemize}{\begin{itemize} \itemsep
-1.5pt}{\end{itemize}}

\onlineid{1091}

\vgtccategory{Technique or Algorithm}

\vgtcinsertpkg

\title{\lexararf: Reference-Free Metrics for Evaluating Conversational Visual Analytics Agents\vspace*{-5pt}}

\author{Srishti Palani\thanks{E-mails: \{srishti.palani, vsetlur\}@salesforce.com.} \and Vidya Setlur\footnotemark[1]}
\affiliation{\scriptsize Tableau Research, Palo Alto, CA, USA}

\abstract{
Conversational visual analytics (CVA) agents powered by large language models generate visualizations and natural-language explanations from open-ended queries. Evaluating these multimodal outputs is challenging: curated reference benchmarks are costly to author, cannot comprehensively capture the space of valid responses, and are unavailable in production. Building on the Lexara evaluation framework, we introduce \lexararf, a reference-free set of metrics that scores CVA outputs using only the prompt, data, and model response. We reformulate evaluation as verification: 13 metrics operationalize visualization design theory and Gricean cooperative principles as computable consistency, intent-alignment, and design validity checks. On a human-rated corpus of CVA test-cases, \lexararf achieves alignment comparable to reference-based formulations, outperforms surface-similarity NLG baselines, and localizes structurally grounded failures with high accuracy.

}


\begin{document}

\firstsection{Introduction}

\maketitle

\label{sec:intro}
Advances in large language models (LLMs) have enabled conversational visual analytics (CVA) systems that allow users to generate and refine visualizations through natural language~\cite{setlur2016eviza, setlur2022converse}. While these systems lower barriers to data exploration, evaluating their outputs remains a challenge. CVA agents produce multimodal outputs, i.e., data-grounded visualizations paired with natural language explanations that must be assessed for data fidelity, encoding effectiveness, and alignment with user intent across a conversation. 

Existing evaluation approaches~\cite{nvbench2025, chen2025viseval, palani2026lexara} compare model outputs against curated references (i.e., acceptable responses for a given input). However, the space of valid CVA responses is inherently open-ended: a single query may admit multiple valid visualizations differing in chart type, encoding, or aggregation, making exhaustive coverage infeasible and introducing false negatives~\cite{fu2020quda}. Moreover, constructing and maintaining references is costly, requiring expert effort to enumerate acceptable outputs and update them as data and tasks evolve~\cite{nvbench2025, chen2025viseval}. In production settings, references are often unavailable, as user queries and data continuously shift, preventing evaluation at inference time and scalable monitoring of CVA agent quality~\cite{sculley2015hidden, Breck2017MLTestScore, lam2011empirical}.

To address these limitations, we propose \lexararf, a reference-free extension of the Lexara evaluation framework~\cite{palani2026lexara} that assesses CVA outputs as intrinsic artifacts grounded in the user prompt and underlying data. We reformulate evaluation as verification: each metric tests whether a response satisfies conditions derived from visualization design theory~\cite{mackinlay1986automating, cleveland1984graphical} and conversational principles~\cite{grice1975logic, setlur2022converse}, rather than whether it matches a curated reference output.  We contribute: (1) 13 reference-free metrics spanning \textit{expressiveness}, \textit{effectiveness}, and \textit{conversational alignment}, with full pseudocode and judge prompts; (2) a validation on a human-rated CVA corpus (N = 120) showing parity with reference-based metrics at this sample size and substantially higher alignment than surface-similarity NLG baselines (BLEU, ROUGE-L, BERTScore); and (3) evidence that the metrics localize structurally grounded failure modes while surfacing semantically ambiguous ones as distributed signals, supporting hybrid evaluation. Together, these contributions establish reference-free evaluation as a practical complement to reference-based benchmarks for CVA, enabling continuous integration, production monitoring, and rapid prompt iteration where reference authoring is prohibitive.

\section{Related Work}
\label{sec:related}
This work builds on three research themes: (1)~theoretical foundations for visualization design, (2)~automated visualization evaluation, and (3)~conversational quality in analytical dialogue. 

\subsection{Visualization Design Theory}
A rich body of work distinguishes between visualization expressiveness (encoding all and only the intended facts) and effectiveness (leveraging perceptual channels for accurate interpretation)~\cite{mackinlay1986automating, cleveland1984graphical}. Prior research further formalizes visual encoding channels and their perceptual properties, alongside principles of graphical integrity that emphasize truthful and efficient data representation~\cite{bertin1983semiology, tufte1983visual}. More recent frameworks organize visualization design across levels from data abstraction to interaction, outlining how systems support analytical workflows~\cite{munzner2009nested, shneiderman1996eyes}. These insights have been operationalized in systems such as Show Me and Draco, which generate visualizations using heuristics and constraint-based reasoning~\cite{mackinlay2007showme, moritz2019draco}. Unlike these generative systems, \lexararf operationalizes the same principles for evaluation, transforming them into reference-free metrics that verify whether generated visualizations satisfy established design constraints. 

\subsection{Automated Visualization Evaluation}
Benchmarks such as nvBench~\cite{nvbench2025}, VisEval~\cite{chen2025viseval}, and NLVCorpus~\cite{srinivasan2021nlvcorpus} evaluate models against curated reference outputs, while Lexara~\cite{palani2026lexara} extends this paradigm with graded metrics spanning visualization and natural language quality. More recent work explores richer evaluation pipelines by decomposing assessments across representation levels, training learned critics from human examples, or using latent similarity representations~\cite{podo2024vieva, pan2025visshepherd, liu2025simvecvis, song2024marrying}. However, all ultimately depend on reference visualizations, either explicitly or implicitly via training data. This dependence incurs high annotation costs, penalizes valid alternative responses, and precludes evaluation in production where references are unavailable. Parallel reference-free work in NLP~\cite{liu-etal-2023, manakul2023selfcheckgpt, wang-etal-2020-asking, kryscinski-etal-2020-evaluating, kim2026evalet} evaluates text without references but do not address the multimodal, design-sensitive nature of CVA. \lexararf bridges this gap by grounding reference-free evaluation in data, visualization design principles, and conversational alignment. 

\subsection{Conversational Quality for Analytical Dialogue}
CVA systems also produce natural language explanations that must be evaluated for communication quality. Grice's cooperative principles (Quality, Quantity, Relation, and Manner) provide a foundation for assessing truthfulness, informativeness, relevance, and clarity~\cite{grice1975logic}. Prior work adapts these principles to analytical dialogue, emphasizing factual consistency between text and visualization, context preservation across turns, coherence, and insightfulness~\cite{tory2019dowhatimean, setlur2022converse, clark1991grounding,north2006insight}. These frameworks are inherently reference-free, evaluating responses based on internal consistency and conversational context rather than comparison to expected outputs. However, they have not been systematically operationalized as computable metrics for CVA systems. \lexararf instantiates these principles through verification by checking consistency across text, visualization, data, and conversation history, enabling end-to-end reference-free evaluation. \rr{Unlike learned critics~\cite{pan2025visshepherd, podo2024vieva} and reference-free NLG evaluators~\cite{liu-etal-2023, manakul2023selfcheckgpt, wang-etal-2020-asking, kryscinski-etal-2020-evaluating, kim2026evalet}, which rely on supervision from reference exemplars, \lexararf requires no reference data at training or evaluation time and grounds its metrics directly in visualization theory and conversational principles.}


\section{Lexara-RF: Reference-Free Evaluation Metrics}
\label{sec:metrics}
\begin{table*}[!htbp]
\centering
\scriptsize
\renewcommand{\arraystretch}{1.05}
\begin{tabular}{@{}p{0.6em} p{2.0cm} p{4.5cm} p{5.5cm} p{4.5cm}@{}}
\toprule
\textbf{Test} & \textbf{User Utterance} & \textbf{Model's CVA Response ($V$, $NL$)} & \textbf{Computed \lexararf Metrics} & \textbf{ Interpretability of Metrics} \\
\midrule
1 &
\codewrap{show top 10 customers by profit in 2023} &
\begin{minipage}[t]{\linewidth}\vspace{0pt}\centering
\includegraphics[width=\linewidth,height=1.2in,keepaspectratio]{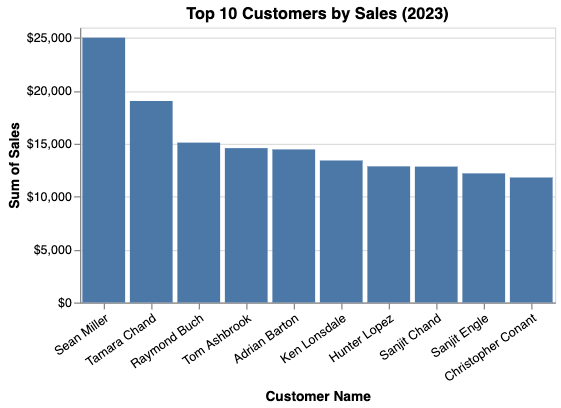}\vspace{2pt}\par
\RaggedRight\scriptsize \textit{``These are the most profitable customers in 2023.''}
\end{minipage} &
\begin{minipage}[t]{\linewidth}\vspace{0pt}\RaggedRight\scriptsize
Data Fidelity = \scorebadge{92\%}; Field Similarity = \scorebadge{50\%}; Sort Accuracy = \scorebadge{70\%}; Filter Accuracy = \scorebadge{100\%}; Chart Type = \scorebadge{100\%}; Axis Accuracy = \scorebadge{100\%}; Visual Encoding = \scorebadge{100\%}; Interactivity = \scorebadge{100\%}. \\ \textbf{Overall Viz} \scorebadge{89\%}.\\
Factual Grounding = \scorebadge{30\%}; Assumptions Disclosure = \scorebadge{25\%}; Insightfulness = \scorebadge{40\%}; Coherence = \scorebadge{90\%}. \\ \textbf{Overall NL} \scorebadge{46\%}.
\end{minipage} &
\begin{minipage}[t]{\linewidth}\vspace{0pt}\RaggedRight\scriptsize
\textbf{Expressiveness and Conversational Alignment failures: field swap.} $V$ picks
\texttt{Sales} where $U$ asked for \texttt{Profit} affecting \emph{Field
Similarity}, \emph{Sort Accuracy}, and \emph{Data Fidelity} scores. \emph{Conversational
Alignment} surfaces the downstream consequence: \emph{Factual
Grounding} and \emph{Insightfulness} drop because $NL$'s ``most profitable'' claim is
incorrect; \emph{Assumptions Disclosure} drops because the silent
\texttt{Profit} to \texttt{Sales} substitution is never
flagged.
\end{minipage} \\[3pt]
2 &
\codewrap{show profit by state} &
\begin{minipage}[t]{\linewidth}\vspace{0pt}\centering
\includegraphics[width=\linewidth,height=1.2in,keepaspectratio]{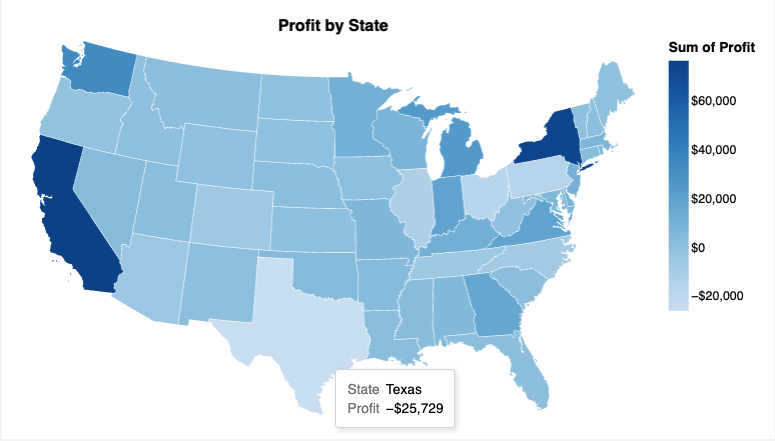}\vspace{2pt}\par
\RaggedRight\scriptsize \textit{``Profit varies across the United States, with California leading the way.''}
\end{minipage} &
\begin{minipage}[t]{\linewidth}\vspace{0pt}\RaggedRight\scriptsize
Data Fidelity = \scorebadge{100\%}; Field Similarity = \scorebadge{100\%}; Filter Accuracy = \scorebadge{100\%}; Sort Accuracy = \scorebadge{100\%}; Chart Type = \scorebadge{100\%}; Axis Accuracy = \scorebadge{100\%}; Visual Encoding = \scorebadge{40\%}; Interactivity = \scorebadge{100\%}. \\ \textbf{Overall Viz} \scorebadge{93\%}.\\
Factual Grounding = \scorebadge{70\%}; Assumptions Disclosure = \scorebadge{50\%}; Insightfulness = \scorebadge{30\%}; Coherence = \scorebadge{90\%}.  \\ \textbf{Overall NL} \scorebadge{60\%}.
\end{minipage} &
\begin{minipage}[t]{\linewidth}\vspace{0pt}\RaggedRight\scriptsize
\textbf{Effectiveness and Conversation Alignment failures:
encoding mismatch.} Sequential single-hue palette on signed
\texttt{Profit} (range $-\$25$K to \$$76$K) renders profit losses visually indistinguishable from low-profit states, affecting \emph{Visual
Encoding}. \emph{Insightfulness}, \emph{Assumptions Disclosure} drop because this obscuring palette choice is not articulated. \emph{Expressiveness} encodes
correct fields and marks.
\end{minipage} \\
\bottomrule
\end{tabular}
\caption{Worked examples showing how \lexararf reference-free metrics
evaluate CVA responses and help diagnose failures.}
\label{tab:reference-free-computation}
\vspace{-4mm}
\end{table*}

We propose \lexararf, a set of reference-free metrics that evaluate CVA outputs as intrinsic artifacts grounded in the user prompt and underlying data source, rather than curated references. \lexararf operates on the following runtime inputs: the user utterance $U$, visualization specification $V$ and natural language explanation $NL$, underlying data source and schema $(D, S)$, and conversation history $H$. While aligned with prior CVA evaluation dimensions, our metrics derive signals directly from visualization theory and conversational principles instead of curated references. They are implemented as verification procedures that test consistency, intent alignment, and design validity, producing graded scores on a common 0--100 scale. Full pseudocode for each metric and additional worked examples spanning aggregation, filter, sort, encoding, and multi-turn failure modes are provided in the paper's Supplementary Materials. \rr{For reproducibility, all implementation choices were fixed a priori: semantic similarity uses \texttt{sentence-transformers/all-MiniLM-L6-v2} embeddings with cosine similarity~\cite{reimers2019sentencebert,wang2020minilm}; fuzzy string matches use \texttt{RapidFuzz} token-set ratios normalized to $[0,1]$~\cite{bachmann2025rapidfuzz}; metric outputs are normalized to the shared $[0,100]$ scale; thresholds include $0.3$ for Data Fidelity plausibility, $\geq 0.5$ for filter field/value matching, and $1.5 \times \mathrm{IQR}$ for outlier detection; and full 13-metric evaluation requires approximately 18 seconds per response, or about USD 5 in API charges for the 120-response corpus based on the OpenAI API pricing schedule~\cite{openai2026pricing}, with details in the Supplementary Materials.}

We employ fixed weights and penalties while scoring the metrics to reflect the relative importance and reliability of different signals, grounded in visualization theory and verification strength rather than learned from data. \rr{The weights, 13 metric dimensions, and 0--100 scoring scale are inherited from Lexara~\cite{palani2026lexara}, preserving interpretability and enabling direct comparison between reference-based and reference-free formulations (§4.2).} 
Signals verified directly against the data receive the highest weight as objective evidence; structural and semantic checks receive moderate weight; multiplicative penalties capture errors with compounding perceptual impact (e.g., axis misuse). 

We organize our reference-free metrics into three themes: \textit{expressiveness}, \textit{effectiveness}, and \textit{conversational alignment}. Rather than comparing responses to references, the metrics verify consistency with visualization theory and conversational principles. 

\subsection{Expressiveness of the Visualization}
Expressiveness captures whether a visualization encodes the intended data and analytical intent. We verify this by checking that the specification $V$ is consistent with the data $(D,S)$ and aligned with the user query $U$. 

\noindent\textbf{Data Fidelity} verifies that the chart's values are computationally correct. We compute three signals: (1) \textbf{$s_1$} (structural validity) checks that fields exist in the schema, operations match data types, and fields plausibly correspond to concepts extracted from $U$; (2) \textbf{$s_2$} (pipeline re-execution) independently executes $V$'s transformation pipeline on $D$ and compares the result to the chart output, treating the specification as the claim and the data as the evidence. Alternative aggregations (e.g., \texttt{SUM} vs. \texttt{COUNT}) are evaluated to detect mislabeling; (3) \textbf{$s_3$} (intent alignment) verifies that the fields and aggregations in $V$ match the analytical intent inferred from $U$. If $s_1 < 50$, the specification is structurally invalid and the score is $s_1$; otherwise, the score is $0.2\,s_1 + 0.5\,s_2 + 0.3\,s_3$, weighting re-execution highest as the strongest correctness signal.

\noindent\textbf{Field Similarity} verifies that the visualization binds the fields implied by the user's query. We extract analytical concepts from $U$ (e.g., ``revenue,'' ``category,'' ``last quarter'') and match each concept to schema columns in $S$. For each concept, we score candidate fields using a weighted combination of semantic similarity, fuzzy string matching, and type compatibility, selecting the highest-scoring column as the best match. The final score measures how many of these best-matched fields are actually used in $V$. This approach captures semantic alignment rather than exact matching: a query for ``\textit{revenue by category}'' should map to fields like \texttt{Sales} and \texttt{Category}, while a chart using unrelated fields such as \texttt{Discount} or \texttt{Region} scores low despite being syntactically valid.

\noindent\textbf{Filter Accuracy} verifies that the visualization applies the constraints implied by the user's query. We extract temporal, categorical, and numeric filter intents from $U$, augmenting relevant filters from prior turns using conversation history $H$. We normalize $V$'s filters and match them to these intents using semantic similarity over fields and values. Scoring uses an asymmetric penalty that prioritizes missing constraints: $100 \cdot m / |F_U| - 25 \cdot \text{miss} - 15 \cdot \text{extra}$, where missing filters are penalized more than unrequested additions. This reflects that omitted constraints violate user intent, while extra filters only narrow the result. Only explicit and persistent filters are evaluated; implicit data source defaults are ignored. 

\noindent\textbf{Sort Accuracy} verifies that the visualization respects ordering implied by the user's query. We detect sort signals in $U$, including superlatives (e.g., ``top,'' ``highest'') and explicit directives (e.g., ``sort by X''), and infer the sort field and direction. If no sort is requested, the metric returns $100$ when $V$ is unsorted. Otherwise, we compare the inferred sort to $V$'s specification. Scoring combines field and direction alignment: $100 \cdot s_f \cdot d$, where $s_f$ measures similarity between expected and actual sort fields, and $d \in {1.0, 0.5}$ rewards correct or reversed direction. Missing required sorting scores $0$.

\subsection{Effectiveness of the Visualization}
Effectiveness captures whether a visualization uses perceptual channels appropriately for accurate interpretation, matching encodings to the data and analytical task~\cite{mackinlay1986automating, cleveland1984graphical, bertin1983semiology}. 

\noindent\textbf{Chart Type} verifies that the chosen mark type aligns with the user's intent and data characteristics. We combine two signals: (1)  intent inferred from $U$ (e.g., trend, comparison, distribution, correlation) to identify expected chart types, and (2) Show Me~\cite{mackinlay2007showme}, which produces a ranked set of appropriate marks from the inferred fields. The score is 100 if $V$ matches the top recommendation, 50 if it appears in the candidate set, and 0 otherwise, with a small bonus for matching the inferred intent. 

\noindent\textbf{Axis Accuracy} verifies that axis assignments support accurate interpretation. Since position is the most precise perceptual channel~\cite{cleveland1984graphical}, incorrect assignments can distort meaning. We audit key conventions with multiplicative penalties: (a) temporal intent requires encoding time on a positional channel; (b) time should appear on the x-axis when both axes are present; (c) bar charts should place measures on the y-axis and categories on x; (d) bar and area charts should use a zero baseline; and (e) explicit axis directives in $U$ must be respected. Violations incur multiplicative penalties to reflect their compounding effect on interpretability.

\noindent\textbf{Visual Encoding} verifies that non-positional channels (color, shape, size, opacity, text) are used appropriately. Perceptual principles recommend matching channels to data types (e.g., hue for categories, size for magnitude)~\cite{bertin1983semiology, moritz2019draco}. We score each channel using \emph{practice} (alignment between channel and data type), and \emph{relevance} (whether the encoded field is implied by $U$). The final score is the mean of $0.6 \cdot$ practice and $0.4 \cdot$ relevance across used channels. Mismatched encodings reduce interpretability, while irrelevant fields introduce visual clutter.

\noindent\textbf{Interactivity} verifies that tooltips provide relevant details on demand~\cite{shneiderman1996eyes}. Required fields include those referenced in $U$ and those defining the visualization (e.g., axes, color). We compute two signals: \emph{coverage}, the fraction of required fields present, and \emph{usability}, which penalizes duplicate fields and missing units. The final score is $0.7 \cdot$ coverage $+ 0.3 \cdot$ usability, capturing both completeness and clarity. 

\subsection{Conversational Alignment}
CVA responses include natural language explanations that must support analytical reasoning and interpretation. We ground evaluation in Grice’s cooperative principles, adapted for analytical conversation, which assess utterances in context rather than against references~\cite{grice1975logic, setlur2022converse}.

\noindent\textbf{Factual Grounding} (Quality) verifies claims in $NL$ against $V$ and recomputed values from $D$, penalizing contradictions more heavily than omissions.

\noindent\textbf{Assumptions Disclosure} (Quantity) checks whether implicit choices in $V$ (e.g., mappings, filters, aggregations) are surfaced in $NL$. Programmatically extracted assumptions are scored by an LLM judge, since undisclosed assumptions cannot enter shared context~\cite{clark1991grounding}.

\noindent\textbf{Insightfulness} extends Quantity by assessing whether $NL$ surfaces meaningful patterns (e.g., trends, outliers), grounded in summary statistics from $D$ and $V$~\cite{north2006insight}.

\noindent\textbf{Coherence} (Manner) evaluates clarity and logical structure of $NL$.

\noindent\textbf{Follow-up Relevance} (Relation) assesses multi-turn alignment using conversation history $H$, verifying persistence of filters, analytical scope, and entity references~\cite{clark1991grounding}.

\noindent \textbf{Overall Viz} and \textbf{Overall NL} are the mean of their constituent metrics, with per-metric scores reported for interpretability.

\section{Validation of Reference-Free CVA Metrics}
\label{sec:validation}
We evaluate \lexararf along three dimensions: \textit{Concurrent
Validity}: alignment with expert human judgments;
\textit{Convergent Validity}: agreement with reference-based
evaluation despite not requiring curated expected outputs, and
discriminant separation from surface-similarity NLG baselines; and
\textit{Diagnostic Utility}: localization of interpretable failure
modes via per-metric score decomposition. All evaluations reuse the human-rated CVA response corpus from
Palani \& Setlur~\cite{palani2026lexara}: $N=120$ responses
stratified by metric score ranges,
ambiguity types (syntactic, semantic, pragmatic), and analytical
task types (descriptive, comparative, trend), drawn from a
generative corpus of evaluation experiments across eight LLMs
(GPT-5 family, o3, o4-mini, Claude Opus 4, Claude 3.7 Sonnet, 
DeepSeek R1), six system prompts, and analyst-supplied data sources
spanning finance, education and healthcare domains.
Each response carries expert human ratings on a 0--100 scale per
metric from two Lexara-trained CVA practitioners (median linear-weighted Cohen's $\kappa = 0.65$ for viz metrics, $0.63$ for NL
metrics, per~\cite{palani2026lexara}). Because of
 the sample size, we present this as a descriptive validation rather than an equivalence test; per-metric results are provided in the Supplementary Materials. 

\subsection{Concurrent Validity: Human Alignment} 
\label{sec:concurrent}
All metrics are computed solely from
$(U, V, NL, D, S, H)$ without reference outputs. Human scores are the mean of two expert ratings. Inter-rater agreement was substantial, with a median linear-weighted Cohen's $\kappa = 0.78$ (IQR $0.71$--$0.83$) across metrics and corpus-wide $\kappa = 0.76$, comparable to prior CVA evaluation on the same corpus~\cite{palani2026lexara}.

Per-metric correlations are reported in the Supplementary Materials, with aggregate values of
Spearman's $\rho_\text{med} = 0.60$ (range $0.42$--$0.74$ across metrics) and \rr{Kendall's $\tau_b{}_\text{med} = 0.47$ (across metrics)}. Deterministic, execution-grounded metrics (Data Fidelity, Filter Accuracy, Sort Accuracy, Factual Grounding) show the strongest alignment with human judgment ($\rho = 0.68$--$0.74$), whereas rubric-based LLM-as-Judge metrics (Insightfulness, Coherence) exhibit lower but consistent agreement ($\rho = 0.42$--$0.52$), reflecting the greater subjectivity of interpretive evaluation~\cite{north2006insight, palani2026lexara}. This distinction also provides practical guidance on which metrics are suitable for automated gating versus human review.


\subsection{Convergent Validity: Reference-Based Alignment}
\label{sec:convergent}

We compare each reference-free metric against (i) its
reference-based counterpart from Palani \&
Setlur~\cite{palani2026lexara} and (ii) standard NLG baselines
(BLEU~\cite{papineni2002bleu}, ROUGE-L~\cite{lin2004rouge},
BERTScore~\cite{zhang2019bertscore}), all evaluated on the same
response set. Agreement with human judgment is assessed using
Steiger's $Z$-test~\cite{steiger1980tests} for dependent
correlations, with Holm--Bonferroni correction across the 13
comparisons.

Reference-free metrics fail to reject parity with reference-based formulations on $11$ of $13$ dimensions ($p > 0.05$ after correction), with mean $|\Delta\rho| = 0.04 \pm 0.01$ on the remaining two dimensions. Because $N=120$ provides limited power to detect small effects, this result should be interpreted as indistinguishable performance at the current sample size, not strict measurement equivalence; detecting $\Delta\rho=0.05$ with 80\% power would require approximately $N\approx400$. These findings support reference-free metrics as a practical complement to reference-based evaluation when reference authoring is prohibitively expensive.

\noindent\textbf{NLG baselines.} BLEU, ROUGE-L, and BERTScore correlate with human judgment at $\rho = 0.18 \pm 0.02$, $0.24 \pm 0.03$, and
$0.46 \pm 0.03$, respectively; substantially below  reference-free and reference-based formulations (see Supplementary Materials). This indicates that surface similarity to a reference $NL$ string does not capture the structurally grounded quality of CVA responses, and that the parity result above is non-trivial: any metric correlating at $\rho \approx 0.60$ with humans would already be doing more than a strong text-similarity baseline.

\noindent\textbf{Divergent case analysis.} We analyzed cases where reference-free and reference-based scores differed by more than 0.15 (90th percentile of pilot-phase score variation), yielding 36 responses. Two annotators independently coded each case (Cohen's $\kappa=0.74$) as valid alternatives, phrasing-invariant disclosures, or true errors, resolving disagreements through discussion~\cite{palani2026lexara, setlur2022converse}. The distribution was 44\%, 32\%, and 24\%, respectively. Although descriptive of this subset only, the low proportion of true errors suggests that most divergences reflect legitimate under-specification in CVA tasks rather than failures of the reference-free metrics. 

\subsection{Diagnostic Utility: Failure Localization}
\label{sec:diagnostic}
To examine whether individual metrics surface their target failure modes, we analyzed a stratified subset of $N=90$ responses from the Lexara corpus~\cite{palani2026lexara} \rr{across performance terciles (high, medium, low) and the three metric pillars. Per-mode counts range from 5 to 11 cases (Supp C Table 2) and reflect the natural distribution of failure types in the source corpus rather than enforced balance; we treat the structurally grounded subset (54 responses across six modes) as our primary basis for localization claims. Two Lexara-trained annotators independently labeled each response using an 11-category failure taxonomy adapted from Lexara (Cohen's $\kappa=0.79$), with disagreements resolved through discussion (Supplementary Materials)}. The taxonomy was derived independently of the metric implementations in Section~\ref{sec:metrics} and comprised:
\begin{tight_itemize}\itemsep0pt
  \item \textbf{Expressiveness:} incorrect data values, field mapping, aggregation, filters, and sorting.
  \item \textbf{Effectiveness:} inappropriate chart type, axis errors, and visual encoding mismatches.
  \item \textbf{Conversational Alignment}: factual grounding failures, undisclosed implicit assumptions, and multi-turn context loss.
\end{tight_itemize}

For each labeled response, we identify the metric with the largest negative standardized deviation from its corpus-level distribution and treat it as the primary diagnostic signal, while assessing whether deviations are localized or distributed. Two examples illustrate this behavior. T47 omitted a year filter from the preceding turn produces a sharp drop in Filter Accuracy while other metrics remain near baseline, cleanly isolating the failure. The failure to disclose an implicit population-normalization assumption in T63 produces coordinated declines in Assumptions Disclosure, Insightfulness, and Coherence, reflecting a conversational failure.

Across the corpus, structurally grounded failures (e.g., missing filters, aggregation errors, baseline violations) produce localized deviations in their corresponding metrics, whereas semantically ambiguous failures (e.g., undisclosed assumptions, pragmatic interpretation) produce correlated declines across conversational alignment metrics. Thus, the framework localizes well-defined structural errors while surfacing ambiguity as distributed signals. Per-failure-mode breakdown in the Supplementary Materials.



\section{Discussion: Limitations \& Future Work}
\label{sec:discussion}
These results support a measured claim: \lexararf aligns with expert judgment at correlations comparable to reference-based methods (median $\rho = 0.60$), exceeds surface-similarity NLG baselines, and localizes structurally grounded failure modes at $\geq 90\%$ accuracy without curated references. At $N=120$, observed parity reflects limited statistical power rather than strict equivalence; accordingly, we position \lexararf as a complement to reference-based evaluation that eliminates per-test-case ground-truth authoring. Rubric-based metrics show lower agreement and weaker localization, warranting different gating thresholds and continued human review. \rr{Three factors plausibly explain why rubric-based metrics underperform: (i) inherent subjectivity in interpretive tasks like Insightfulness and Coherence~\cite{north2006insight}; (ii) LLM-as-Judge variance even at temperature 0 (details in supplementary materials); and (iii) scope overlap among Insightfulness, Coherence, and Assumptions Disclosure, which manifests as correlated rather than localized signals when failures are semantically ambiguous (§4.3). Promising improvement paths include ensemble judging across model families, calibration of rubric anchors against held-out human ratings, and decomposing broader rubrics into narrower verification primitives.} Only 24\% of reference-free vs. reference-based disagreements (\S\ref{sec:convergent}) are true metric errors, whereas 44\% are \emph{valid alternatives}, suggesting much of what reference-based evaluation flags is legitimate under-specification rather than metric failure. \rr{Per-metric decomposition effectively localized failure dimensions in our case studies, suggesting promise for regression triage and prompt debugging in CVA pipelines.}

\rr{Three limitations bound our claims: First, the framework is validated only on Vega-Lite-style specifications; extending metrics that inspect encodings or interaction requires adaptation to other grammars (e.g., ggplot, D3). Second, evaluation is limited to English utterances and judge prompts. Third, validation uses the Lexara corpus with inherited metric weights; broader datasets spanning grammars, languages, and analytical tasks are needed for data-driven calibration.} Future work should target a larger validation corpus spanning multiple grammars and languages, data-driven weight calibration, and extension to richer CVA tasks such as multi-table joins and calculated fields. Our results highlight a hybrid evaluation paradigm: automated, reference-free metrics like \lexararf can reliably handle structurally grounded correctness checks, while human judgment remains essential for interpretive and subjective aspects of analytical communication.

\rr{\textbf{Data Availability.} The 120-response validation corpus is available at \url{https://github.com/SrishtiPalani/Lexara-CVA-Eval/tree/main/lexara_rf_corpus}. Supplementary materials have all details for reproducibility. }

\newpage
\bibliographystyle{abbrv-doi}
\bibliography{main}

\end{document}